\newcommand{\be}{\begin{equation}}
\newcommand{\ee}{\end{equation}}
\documentclass[twocolumn,a4paper,10pt]{article}
\usepackage[dvips]{graphicx}

\title{Universality in snowflake aggregation}
\author{C. D. Westbrook, R. C. Ball, P. R. Field \& A. J. Heymsfield}
\begin{document}

\maketitle

\begin{abstract}
Aggregation of ice crystals is a key process governing precipitation.
Individual ice crystals exhibit considerable diversity of shape, and
a wide range of physical processes could influence their aggregation;
despite this we show that a simple computer model captures key features
of aggregate shape and size distribution reported recently from Cirrus clouds. The results prompt a new way to plot
the experimental size distributions leading to remarkably good dynamical
scaling. That scaling independently confirms that there is a single
dominant aggregation mechanism at play, albeit our model (based on
undeflected trajectories to contact) does not capture its form exactly.
\end{abstract}

\section{Introduction}

It has recently become possible to collect large samples of high resolution
cloud particle images in real time, opening up the modelling of cloud
dynamics to detailed comparison with nature. Figure 1
shows ice crystal aggregates from a Cirrus cloud over the USA, captured
by non-contact aircraft-based imaging.  Such aggregates can be seen
to be comprised of varied rosette ice crystal types, and detailed statistics have recently been published on both the cluster aspect ratios [{\it Korolev and Isaac} 2003]  and size distributions [{\it Field and Heymsfield} 2003] in Cirrus clouds.  Such aggregation
is a key feature of cloud development in the troposphere and can be
quite crucial to the development of precipitation, whether it reaches
the ground as snow or melts first to arrive as rain.

The openness of the aggregates significantly accelerates their growth.
Two clusters (labelled by $i,j$) pass approach with their centres
closer than the sum of their radii $r_{i}+r_{j}$ at a rate proportional
to\begin{equation}
\pi \left(r_{i}+r_{j}\right)^{2}\left|v_{i}-v_{j}\right|,\label{eq:kernel}\end{equation}
 where for each cluster the sedimentation speed $v$ varies inversely
with its radius $r$ and mass $m$ as \begin{equation}
v=\frac{\eta /\rho }{r}F\left(\frac{mg\rho }{\eta ^{2}}\right).\label{eq:fallspeed}\end{equation}
Here $\eta $ and $\rho $ are the viscosity and density of the
air, $g$ is the acceleration due to gravity, and we have assumed that only one geometrical radius is relevant:  {\it Mitchell} [1996] discusses an elaboration. Given the above, the rates of aggregation per unit time for fixed cluster
masses vary linearly overall
with the cluster radii, and openness of aggregate structure enhances
aggregation rates despite lowering fall speed. For real aggregates
this is a significant factor: using data from {\it Heymsfield et al.}, [2002],
one finds that rosette aggregates 2mm across (which yield 0.5 mm droplets)
aggregate four times faster than when melted. For cloud particles
it is also relevant to consider the rates of aggregation per unit
of distance fallen (rather than per unit time), which at fixed mass
is proportional to the square of radius, leading to 16 times enhancement
for ice over water in the example cited.

We have made computer simulations of ice aggregation based on equations
(\ref{eq:kernel}) and (\ref{eq:fallspeed}), tracing trajectories
through possible collisions to obtain accurate collision geometries.
We assumed that all collisions led to rigid irreversible joining of
clusters, as the openness of the experimentally observed clusters
suggests little large scale consolidation of structure upon aggregation,
and that cluster orientations were randomised in between collisions
but did not change significantly during them. We took the sedimentation
speeds to be governed by inertial flow, for which the mass dependence
function in equation (\ref{eq:fallspeed}) is given by $F(X)\propto X^{1/2}$.
Details of implementation are given in a longer paper [{\it Westbrook et al.}, 2004].

Some representative computer aggregates are shown in figure 1 alongside
the experimental ones. Our simulations used three dimensional cross
shapes for the initial particles as a crude representation of the
experimental bullet rosettes.%

Figure 2 shows a quantitative comparison
of aggregate geometry, in terms of the ratio of cluster spans perpendicular
to and along the direction of maximal span, as measured from projections
in an arbitrary plane of observation. We find that different initial
particle geometries (rosettes, rods) approach a common asymptotic
average cluster value. The aspect ratio of CPI images have been similarly
calculated [{\it Korolev and Isaac} 2003], the results of which have been
overlayed onto figure 2, and these appear
to approach the same value. This universality of aspect ratios provides
direct support for our hypothesis of rigid cluster joining upon contact.

A deeper indicator of universality is provided by the fractal scaling
of ice crystal aggregates, where one tests the relation $m\propto \ell^{d_{f}}$ between aggregate mass $m$ and linear span $\ell$.
Our simulations and experimental observations [{\it Heymsfield et al.}, 2002],
rather accurately agree on the fractal dimension $d_{f}=2.05\pm 0.1$
and in {\it Westbrook et al.} [2004] we discuss theoretical arguments
leading to $d_{f}=2$.

Our simulations conform well to dynamical scaling of the cluster size
distribution. This means that number of clusters per unit mass varies
with mass and time together as a single scaling function,\begin{equation}
\frac{dN}{dm}(m,t)=S(t)^{-2}\Phi \left(\frac{m}{S(t)}\right),\label{eq:dyn.scalg}\end{equation}
 where $S(t)$ is the weight average cluster mass. This relationship
is confirmed in figure 3, where we rescale
the mass distribution from different times in the simulations onto
a universal curve.

The scaling function which we observe in figure 3
exhibits power law behaviour with $\Phi (X)\propto X^{-\tau }$ for
$X\ll 1$ with $\tau \approx 1.6$. This is not intrinsically surprising
(and indeed it matches theoretical expectations [{\it Van Dongen and Ernst}, 1985])
but it has forced us to abandon the way experimentally observed distributions
of cluster linear size have hitherto been plotted. The problem is
that given equation (\ref{eq:dyn.scalg}) and its observed power law
form, we must expect that the distribution of clusters by linear span
$\ell $ should at small $\ell $ take the form $\frac{dN}{d\ell }(\ell ,t)\propto \ell ^{-1-(\tau -1)d_{f}}$
which diverges as $\ell ^{-2.2}$ using our observed exponents. For
small enough crystal sizes this behaviour will be modified by the
role of growth processes other than aggregation, but that lies outside
the scaling regime.

Because of the divergence one has to take great care in constructing
a characteristic linear size $L(t)$, where the natural choices are
$L_{k}(t)=\frac{\sum _{clusters}\ell ^{k+1}}{\sum _{clusters}\ell ^{k}}$
and the lowest whole number $k$ for which the denominator is not
dominated by the smallest clusters is $k=2$. The simplest natural
scaling ansatz for the cluster span distribution is then found [{\it Westbrook et al.},  2004]
to be\[
\frac{dN}{d\ell }(\ell ,t)=M_{2}(t)\: L_{2}(t)^{-3}\Psi \left(\frac{\ell }{L_{2}(t)}\right),\]
 where $M_{k}(t)=\sum _{clusters}\ell ^{k}$. Figure 4 shows that
this scaling ansatz works acceptably for our simulation data and well
for the experimental observations. The latter are rich data because
cluster span is one of the simplest automated measurements to take.

The rescaled distributions from simulation and experiment agree fairly
well but not perfectly, as shown in Figure 4.
One experimental reservation is the fall-off of experimental observation
efficiency at small sizes, where clusters can be missed. However our
scaling procedure itself is in effect expressly designed to avoid
sensitivity to this, and the superposition of the experimental data
down to small reduced sizes looks good. Indeed it looks so good that
the transient flattening around $\ell /L\approx 1$, which is absent
from the simulations, appears to be significant.

One suggestion for the flattening around the middle of the rescaled
distribution is that it might be associated with the peculiar feature
of the collision rate being zero between clusters of equal sedimentation
speed. Our simulations include this feature but for low relative approach
speeds, where each cluster has more time to adjust its momentum in
response to the other perturbing the local airflow, our approximation
of ignoring hydrodynamic interactions (and hence phenomena such as
wake capture) is less accurate.

In summary, we have a fairly complete understanding of the geometry
of the atmospheric ice crystal aggregates, dominated by sticking upon
encounter. Further details of the sticking mechanism (which we did
not include) appear not to be important for the cluster geometry,
and the excellent scaling superposition of the experimental cluster
size distributions suggests sticking efficiency does not favour aggregation
at particular sizes. The simplest interpretation of these observations
is that although the sticking probability might be low for a single
microscopic event, many such contacts will be attempted during a cluster-cluster
encounter so that eventual escape is unlikely. The actual sticking
mechanism between ice crystals remains an intriguing open question,
particularly for the low temperatures of figure 1.

The fact that the same evolution is seen for differing initial monomer
populations (rods and rosettes) suggests that a single set of geometric
relationships for ice aggregates can successfully be applied in a
wide range of cloud conditions. This would lead to greater accuracy
in retrieving cloud properties such as precipitation rate and predicting
the radiative affect of ice crystal aggregates upon the climate system.

 \begin{figure}
 \noindent\includegraphics[width=8.4cm]{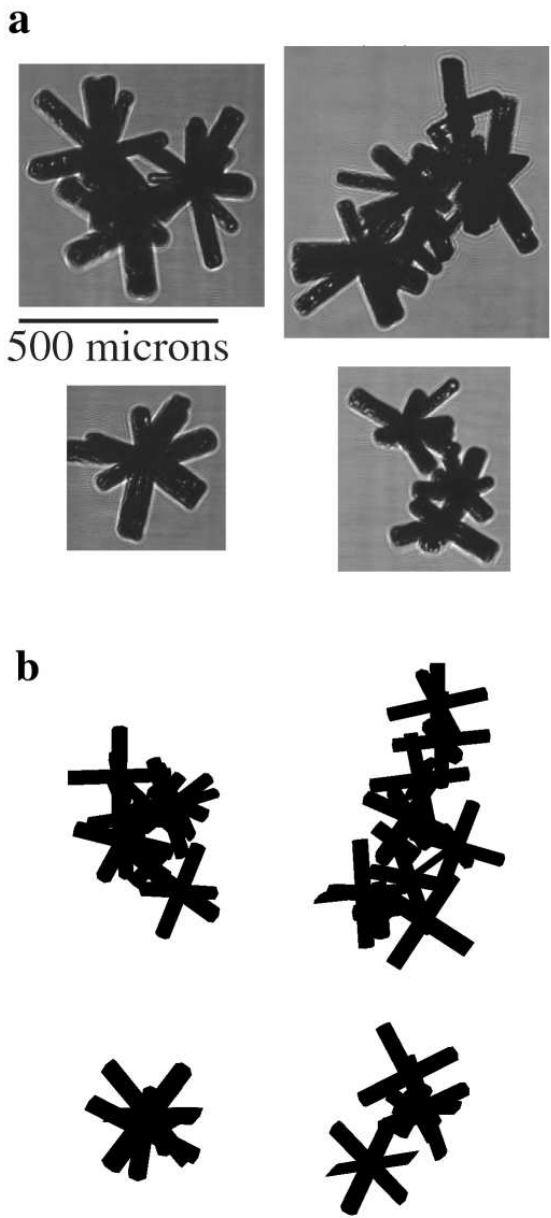}
 \caption{ (a) Ice crystal aggregates images
obtained from an aircraft flight through cirrus cloud, at temperatures
from $-44^{\circ }$C to $-47^{\circ }$C ($\sim $9 km altitude),
using a cloud particle imager (CPI, SPEC Inc., USA). The pictures
shown are aggregates of rosette ice crystal types. (b) Aggregates
as simulated by our computer model which assumed rigid joining when
clusters collide under differential sedimentation. }
 \end{figure}

 \begin{figure}
 \noindent\includegraphics[width=8.4cm]{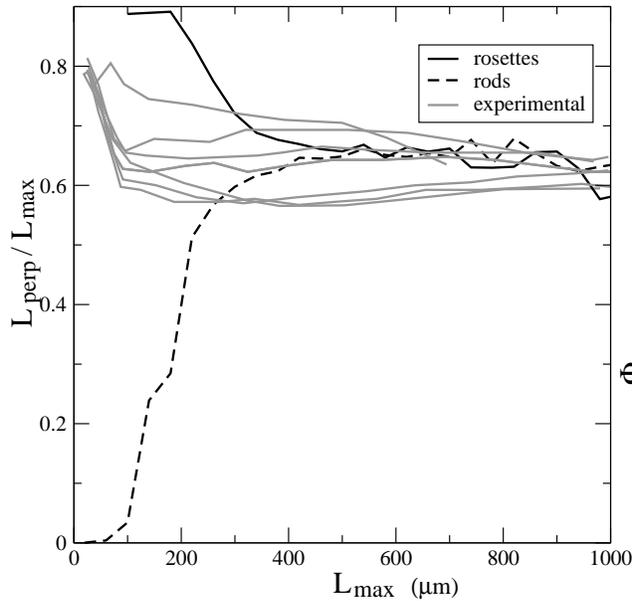}
 \caption{ Mean aspect ratio for projected ice
agggregate images, where the aspect ratio is measured as the longest
span $L_{max}$ divided into the span perpendicular to the longest
$L_{perp}$. Grey lines show cloud data of {\it Korolev and Isaac} [2003] plotted against longest span in microns for a range of temperatures
between $0^{\circ }$C and $-40^{\circ }$C. Black lines show simulation
data plotted against longest span in arbitrarily scaled units, where
the initial particles were three dimensional crosses (solid line)
and simple rods (dashed). }
 \end{figure}

 \begin{figure}
 \noindent\includegraphics[width=8.4cm]{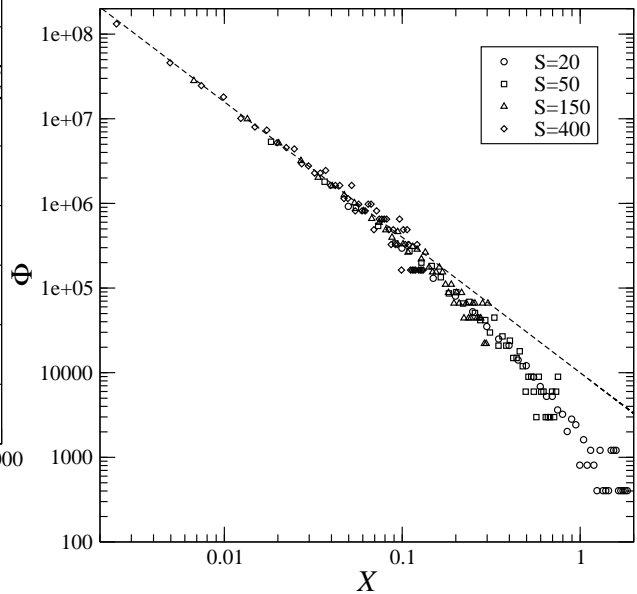}
 \caption{ Dynamical scaling of the cluster mass
distribution from simulations. The superposition of data from different
times supports equation (\ref{eq:dyn.scalg}), and the linear portion
indicates $\Phi (X)\propto X^{-\tau }$ at small $X$ with exponent
$\tau =1.6$. }
 \end{figure}

 \begin{figure}
 \noindent\includegraphics[width=8.4cm]{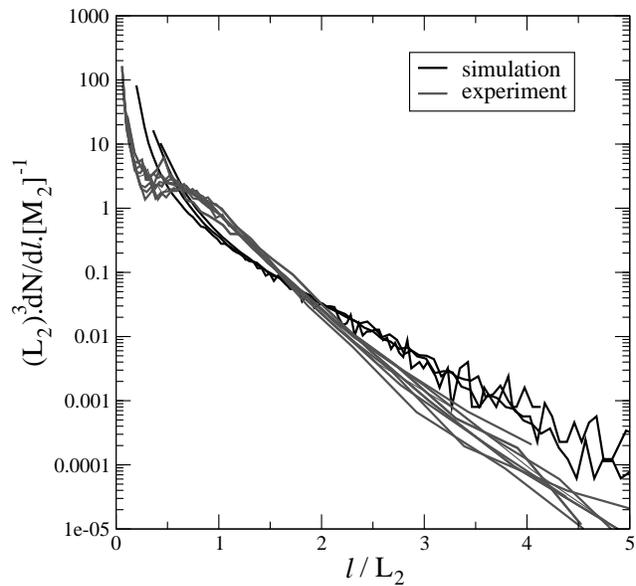}
 \caption{ Cluster length distribution,
rescaled as discussed in the text. The gray lines show experimental
distributions at altitudes of 9.5km ($-50^{\circ }$C) to 6.6km ($-28^{\circ }$C)
in the cirrus cloud of {\it Field and Heymsfield} [2003] obtained
during an ARM (Atmospheric Radiation Measurement program) flight (9th
March 2000). Each experimental size distribution represents an in-cloud
average over 15 km. Black lines show simulation data. }
 \end{figure}

\end{document}